\documentclass[11pt]{article}

\usepackage[margin=1in]{geometry}
\usepackage[T1]{fontenc}
\usepackage{amsmath,amssymb}
\usepackage{graphicx}
\usepackage{booktabs}
\usepackage{xcolor}
\usepackage{multirow}
\usepackage{subcaption}
\usepackage{enumitem}
\usepackage{algorithm}
\usepackage{algpseudocode}
\usepackage{float}
\usepackage[numbers,sort&compress]{natbib}
\usepackage[hidelinks]{hyperref}
\usepackage{microtype}

\graphicspath{{figures/}}
\title{Resource versus Responsiveness: Benchmarking SDN Controller Runtimes for a 
  Moving-Target-Defense Control Plane at Scale}

\author{%
Souhail Chakkour \quad Umesh Biswas \quad Charan Gudla\\[0.4em]
\small Mississippi State University, Mississippi State, USA\\
\small \texttt{sc3130@msstate.edu} \quad
\texttt{ucb5@msstate.edu} \quad
\texttt{gudla@cse.msstate.edu}
}
\date{}

\begin{document}
\maketitle

\begin{abstract}
Network Moving Target Defense (MTD) built on Software-Defined Networking (SDN)
continuously rotates host-facing addresses to invalidate an attacker's
reconnaissance. Such a defense is only as good as the control plane that drives
it: every rotation is a burst of flow mutations, and every new connection is a
reactive flow install that must complete before the address moves again. Yet the
choice of \emph{SDN controller runtime}---the framework that schedules these
mutations---is treated as an implementation detail in the MTD literature. We show
it is not. We port a single, identical MTD controller (CPAM) to three widely used
runtimes---Ryu (single-threaded cooperative Python), OpenDaylight, and ONOS (both
multi-threaded JVM)---and benchmark them under an identical 500-host campus fabric
using an RFC~8456-aligned methodology with ten runs per controller.
All three runtimes deliver the \emph{same} data-plane correctness (near-zero loss,
sub-millisecond jitter) but diverge sharply on the control plane: the two JVM
controllers keep reactive round-trip time low and flat and preserve all sessions
across rotations, whereas Ryu's cooperative scheduler serializes reactive flow
installs behind the periodic rotation burst, inflating reactive RTT by roughly
$100\times$ and dropping a small fraction of connections at setup time. Ryu,
in turn, is markedly lighter, with roughly an order-of-magnitude smaller memory
footprint. Crucially, the two JVM runtimes are \emph{not} interchangeable: ONOS
attains OpenDaylight-class reactive latency at the lowest CPU utilization of the
three and a smaller live heap than OpenDaylight, showing that low reactive latency
need not carry OpenDaylight's full resource cost. The controller runtime thus imposes
a concrete \emph{resource-versus-responsiveness} trade-off on the MTD control plane.
We root-cause each difference to the runtime's concurrency and flow-programming model
and argue that controller selection should be a first-class, measured decision in MTD
system design.
\end{abstract}

\vspace{0.75em}

\section{Introduction}
\label{sec:introduction}

Moving Target Defense (MTD) reduces the value of adversarial
reconnaissance by repeatedly changing the system attributes on which
attackers depend~\cite{jajodia2011mtd,okhravi2013survey,
zhuang2014theory_mtd}. In network-based MTD, a prominent approach is
to periodically change the virtual IP addresses exposed by protected
hosts. Consequently, information obtained through host discovery,
port scanning, and topology inference becomes stale before it can be
reliably used in an attack~\cite{lyon2009nmap,durumeric2013zmap,
carroll2014address_shuffling,alshaer2012rhm,
jafarian2015adversaryaware}.

Software-Defined Networking (SDN) provides a natural platform for
implementing address mutation. A logically centralized controller can
maintain virtual-to-real address mappings, translate packets, and
reprogram the forwarding fabric without modifying end
hosts~\cite{mckeown2008openflow,kreutz2015sdn_survey}. However, the
security benefit of address mutation depends partly on how frequently
addresses change. More frequent mutation shortens the useful lifetime
of reconnaissance information, but also places greater pressure on the
SDN control plane.

Each \emph{mutation interval}---the fixed period between two consecutive
address rotations---ends with a burst of control-plane operations. During
this burst, the controller updates virtual-to-real address mappings and
modifies the corresponding forwarding rules in the switches. Meanwhile,
a newly arriving connection may trigger \emph{reactive flow installation},
in which the controller processes a \texttt{PacketIn} event and installs
the rules needed to forward the connection's packets. Established
connections may also require \emph{retained session state}, meaning
per-session bindings and forwarding rules associated with the virtual
address under which the session was created, so that the session remains
valid after that address rotates. The controller must therefore handle
address mutation, new-flow setup, and session-state maintenance
concurrently. If these operations are delayed or improperly ordered, a
new connection may use an address or rule that is being replaced, creating
a \emph{rotation race} that increases connection-establishment latency or
causes the connection attempt to fail.

Although SDN-based address mutation has been studied extensively,
existing work generally treats the controller framework as an
implementation choice rather than as an experimental
variable~\cite{cho2020proactive,sun2023survey,souto2026mtd}. This
assumption can be misleading because controller runtimes differ
substantially in their concurrency, event-dispatch, state-management,
and flow-programming architectures. A single-threaded cooperative
runtime may serialize reactive processing behind a mutation burst,
whereas a multi-threaded runtime may execute these activities
concurrently at the cost of greater CPU and memory consumption.
General SDN controller benchmarks measure latency, throughput, and
scalability~\cite{rfc8456}, but do not reproduce the bursty and
session-stateful workload created by address mutation. This raises the central question of this paper:
\emph{How does the SDN controller runtime affect the \underline{responsiveness},
\underline{continuity}, and \underline{resource cost} of an address-shuffling MTD when the
underlying MTD logic is held constant?}

To answer this question, we implement
Continuity-Preserving Address Mutation (CPAM)~\cite{chakkour2026balancing} on
Ryu~\cite{ryu2013framework}, a cooperative single-threaded Python controller,
and on two multi-threaded JVM platforms---OpenDaylight
(ODL)~\cite{medved2014opendaylight}, which programs flows through its MD-SAL
datastore, and ONOS~\cite{berde2014onos}, which uses a lighter in-memory flow
subsystem. All three implementations use the same VIP lifecycle, session bindings,
translation placement, rotation parameters, and reclamation rules; the controller
runtime is the principal independent variable.

We evaluate all three controllers over ten runs on an identical three-tier fabric
with 500 hosts and 19 OpenFlow switches using RFC~8456-aligned metrics. All three
provide near-zero UDP loss, sub-millisecond jitter, and preserve established
sessions. The JVM controllers achieve low, flat reactive latency
($0.131\pm0.014$\,ms RTT and a $1.4$\,ms p99 processing time for OpenDaylight;
$0.159\pm0.014$\,ms and $1.2$\,ms for ONOS), compared with $15.98\pm8.41$\,ms and
$15.4$\,ms for Ryu. Ryu instead uses substantially less memory and
achieves a higher aggregate flow-setup rate. The two JVM runtimes are themselves
distinct: ONOS matches OpenDaylight's reactive latency at the lowest CPU of the
three and a smaller heap, so the low-latency operating point does not carry a single
fixed resource cost.\smallskip

\noindent\textbf{Our Contributions.} This paper makes the following contributions:

\begin{itemize}[leftmargin=*,nosep]
    \item We present, to our knowledge, the first controlled comparison
    of SDN controller runtimes for address-shuffling MTD in which the
    MTD logic, topology, workloads, and configuration are held constant.

    \item We develop an RFC~8456-aligned benchmarking methodology
    tailored to address mutation, including rotation-aware reactive
    latency, session continuity, and fair Python-versus-JVM resource
    measurements.

    \item We quantify the resource-versus-responsiveness trade-off
    across Ryu, OpenDaylight, and ONOS on a 500-host campus fabric over ten
    runs per controller, showing that the two JVM runtimes occupy distinct
    operating points rather than a single one.

    \item We provide a root-cause analysis connecting the observed
    latency, continuity, throughput, and resource differences to the
    controllers' cooperative and thread-pooled concurrency models.
\end{itemize}

\section{Background and Motivation}
\label{sec:background}

This section introduces SDN-based address-shuffling MTD, explains the
control-plane workload created by address mutation, and reviews the prior work
that motivates our controller-runtime comparison.

\subsection{SDN-Based Address-Shuffling MTD}

Prior work establishes address mutation as a practical MTD mechanism for
reducing the lifetime of reconnaissance information and increasing adversarial
uncertainty~\cite{jajodia2011mtd,okhravi2013survey,
zhuang2014theory_mtd,carroll2014address_shuffling,alshaer2012rhm,
jafarian2015adversaryaware}. In an SDN-based realization, each protected host
retains a stable real IP address while communicating peers use a short-lived
virtual IP address (VIP). The controller maintains the virtual-to-real
mappings and installs the corresponding forwarding and translation
rules~\cite{10.1145/2342441.2342467,macfarland2015sdn_shuffle,
kreutz2015sdn_survey}.

We define the \emph{mutation interval} as the time between two consecutive VIP
rotations. At each rotation, the controller assigns new primary VIPs and
updates the associated control and forwarding state. Shorter intervals
therefore increase the frequency of control-plane updates, creating the
runtime challenge examined in Section~\ref{sec:controller-challenge}.

Our evaluation adopts the Continuity-Preserving Address Mutation (CPAM) design
introduced by Chakkour et al.~\cite{chakkour2026balancing}. CPAM pins the VIPs used
when a session is established, allowing that session to continue after newer
VIPs are assigned. Both implementations use the same VIP lifecycle, session
bindings, translation placement, rotation parameters, and state-reclamation
rules. The address-mutation mechanism is therefore fixed while the controller
platform is varied.

\subsection{The Controller-Runtime Challenge}
\label{sec:controller-challenge}

Address-shuffling MTD creates a control-plane workload that differs from
conventional reactive SDN forwarding in three important ways.

First, each address rotation creates a \emph{periodic mutation burst}. During
this burst, the controller generates new address mappings, updates translation
state, modifies forwarding rules, publishes the new mappings, and eventually
reclaims retired addresses. These operations are concentrated around each
rotation epoch rather than distributed uniformly over time~\cite{10.1145/2342441.2342467,jafarian2015adversaryaware,
chakkour2026balancing}.

Second, newly arriving connections may require \emph{reactive flow
installation} while a mutation burst is in progress. When a packet does not
match an existing switch rule, the switch sends a \texttt{PacketIn} event to
the controller. The controller must resolve the destination VIP, determine the
forwarding path, and install the required rules before the connection can
proceed~\cite{rfc8456}. If reactive processing is delayed behind mutation work, connection
establishment may experience increased latency.

Third, continuity-preserving mutation requires \emph{retained session state}.
The controller must maintain the address bindings and forwarding rules
associated with established sessions until those sessions terminate. It must
therefore create, access, and reclaim session state while simultaneously
processing new connections and periodic rotations~\cite{chakkour2026balancing,zal2024lossless,11214661}.

These activities can also create a \emph{rotation race}. Such a race occurs
when a new connection is being established while the mapping or forwarding
state associated with its destination VIP is being replaced. Depending on the
ordering and duration of these operations, the connection may be delayed, may
use stale state, or may fail during setup.

How a controller handles these competing activities depends on its runtime
architecture. A cooperative single-threaded runtime may serialize reactive
flow installation behind mutation work. A multi-threaded runtime may overlap
the two activities through separate worker threads, but may require additional
CPU, memory, synchronization, and state-management overhead. Thus, two
controllers executing identical MTD logic may provide the same forwarding
correctness but differ substantially in reactive latency, connection setup,
flow-installation throughput, and resource consumption. Section~\ref{sec:arch}
examines the relevant architectural differences in detail.

\subsection{Existing Practice and Research Gap}

SDN-based address-shuffling MTD has been studied extensively, including
mutation design, adversary-aware address randomization, and transparent
deployment~\cite{carroll2014address_shuffling,alshaer2012rhm,
jafarian2015adversaryaware,macfarland2015sdn_shuffle,
cho2020proactive,sun2023survey}. Jafarian et
al.~\cite{10.1145/2342441.2342467,jafarian2015adversaryaware}
established foundational OpenFlow-based host-mutation mechanisms. Later work
has considered session continuity~\cite{zal2024lossless,11214661},
recognition of active MTD mechanisms~\cite{mtdsense2024}, and deployment in
IoT environments~\cite{sharma2025ttc}. A recent systematic
review~\cite{souto2026mtd} identifies more than 80 SDN-based MTD systems.

This literature primarily evaluates mutation strategies, security
effectiveness, session preservation, and deployment mechanisms. The
controller platform is typically selected as an implementation choice rather
than treated as an experimental variable. Consequently, reported latency,
throughput, or resource overhead may reflect both the MTD mechanism and the
architecture of the controller that executes it.

SDN controller performance has also been studied independently of
MTD~\cite{mendonca2023performance,narantuya2019sdn_multicontroller}.
RFC~8456~\cite{rfc8456} defines metrics including reactive latency,
flow-setup throughput, scalability, and resource consumption. However,
conventional controller benchmarks generally model reactive forwarding
without the periodic mutation bursts, retained session state, and
rotation-overlapping connection setup introduced by address-shuffling MTD.

The two research areas therefore leave a clear gap: MTD evaluations rarely
isolate controller-platform effects, while controller benchmarks do not
reproduce an MTD-specific control-plane workload. We address this gap through
a controlled comparison of Ryu, OpenDaylight, and ONOS in which the CPAM mechanism,
topology, traffic, configuration, and measurement procedure are held constant.
This design isolates how controller architecture affects reactive
responsiveness, session continuity, flow-programming behavior, and resource
consumption under the same address-mutation workload.

\section{Controller Architecture}
\label{sec:arch}

\begin{figure}[t]
\centering
\includegraphics[width=0.8\textwidth]{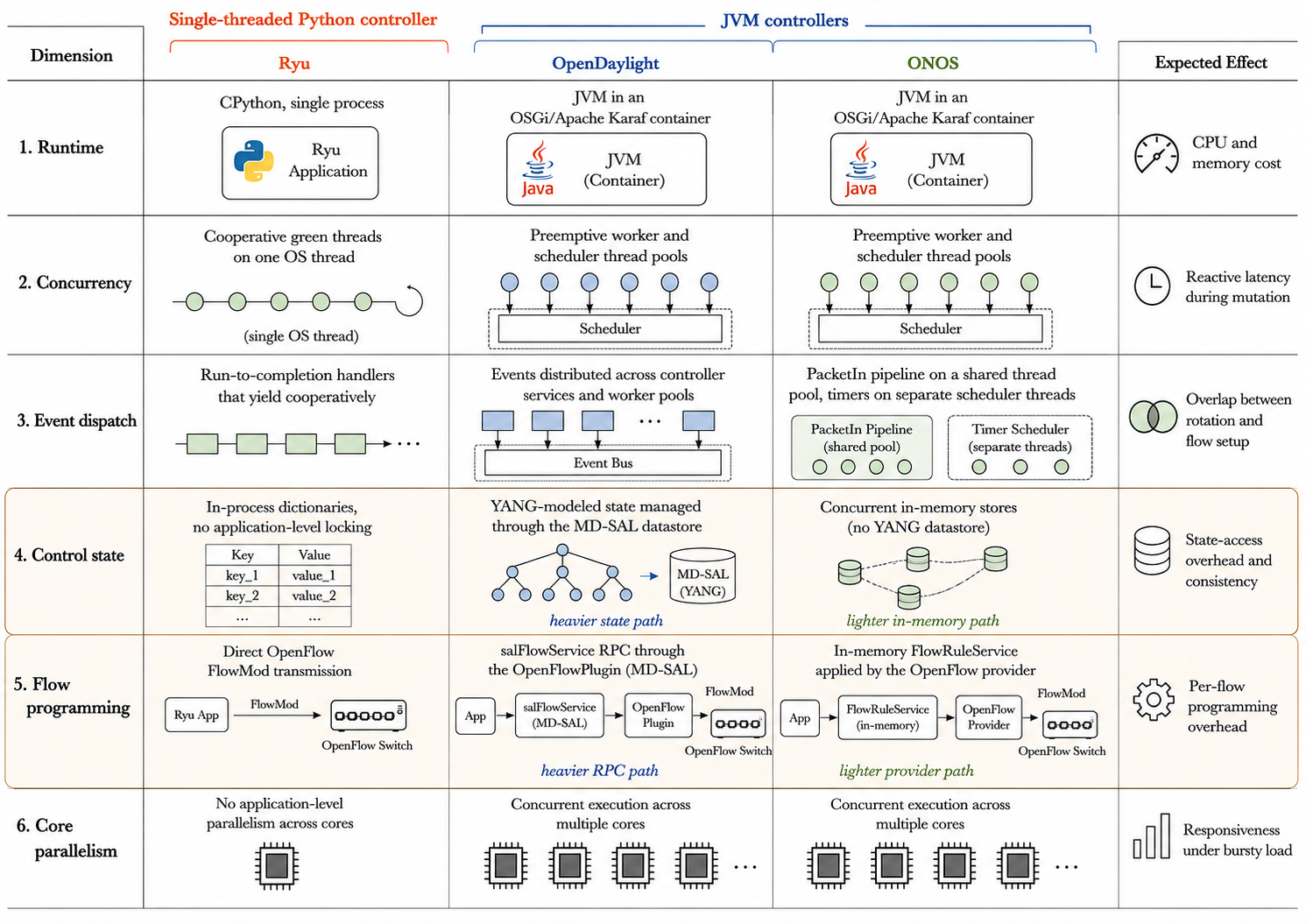}
\caption{Controller architectures under identical CPAM control logic. Ryu executes cooperatively on a 
single thread, while OpenDaylight and ONOS are both multi-threaded JVM runtimes that differ chiefly in 
their control-state and flow-programming paths (MD-SAL datastore vs.\ in-memory FlowRuleService).}
\label{fig:arch}
\end{figure}

All three implementations execute the same CPAM control logic under the same
configuration and workload. The controller runtime is therefore the principal
independent variable in our study. This section compares Ryu, OpenDaylight, and
ONOS along the architectural dimensions most relevant to periodic address mutation:
event scheduling, state management, and flow programming. Ryu is a single-threaded
Python controller; OpenDaylight and ONOS are both multi-threaded JVM platforms, but
they differ in how flows are programmed---OpenDaylight through a YANG/MD-SAL
transactional datastore, ONOS through a lighter in-memory flow subsystem---which, as
the results show, separates them on resource cost. Figure~\ref{fig:arch}
summarizes these differences and their expected effects.

\subsection{Execution and Event Dispatch}

Ryu~\cite{ryu2013framework} is a lightweight Python controller that uses
cooperative green threads. Packet-in processing, periodic timers, address
rotation, and statistics collection share one operating-system thread. A
handler continues until it completes or explicitly yields at an I/O or sleep
operation. This execution model simplifies state access because controller
handlers generally do not modify the mapping tables concurrently. However, a
long-running handler can delay other controller activities.

This behavior is particularly relevant to CPAM. At each mutation interval, the
rotation handler updates the VIP assignments and associated forwarding state
for many hosts. A \texttt{PacketIn} arriving during this work may remain queued
until the rotation handler yields. Ryu cannot use additional processor cores to
execute the reactive handler in parallel with the rotation task.

OpenDaylight~\cite{medved2014opendaylight} runs on a multi-threaded JVM-based
platform. Its OpenFlowPlugin dispatches southbound events through worker
threads, while application timers and service calls are handled through
controller scheduler and service threads. Periodic rotation and reactive flow
installation may therefore proceed concurrently on different cores.

This parallel execution can reduce interference between mutation and
connection setup. It also introduces costs absent from Ryu, including thread
scheduling, synchronization, service abstraction, datastore processing, JVM
heap management, and garbage collection.

ONOS~\cite{berde2014onos} is likewise a multi-threaded JVM platform, built on an
OSGi (Apache Karaf) runtime. Southbound \texttt{PacketIn} events are delivered to a
pipeline of packet processors running on a shared thread pool, while periodic timers
(rotation, reclamation, statistics) execute on separate scheduler threads. Like
OpenDaylight, ONOS can therefore overlap reactive flow installation with an ongoing
rotation on different cores; unlike OpenDaylight, it does not interpose a
transactional model-driven datastore on the flow-programming path (below), which
bears directly on its resource profile.

\subsection{State Management and Flow Programming}

Ryu stores CPAM's host mappings and session bindings in ordinary in-process
Python dictionaries. Because the application executes cooperatively on one
thread, these structures require no application-level synchronization during
normal event handling. Ryu installs forwarding rules by constructing
OpenFlow \texttt{FlowMod} messages and sending them directly through the
switch datapath connection.

OpenDaylight represents the corresponding state through its YANG-based
Model-Driven Service Abstraction Layer (MD-SAL). The OpenDaylight
implementation programs forwarding rules through the
\texttt{salFlowService} interface, which the OpenFlowPlugin translates into
southbound OpenFlow messages. This path provides structured state management
and transactional controller services, but adds software layers between the
application and the switch.

ONOS keeps CPAM's mapping and session state in ordinary concurrent in-memory
structures and programs forwarding rules through its \texttt{FlowRuleService},
whose flow store is applied to the switch by the OpenFlow provider. This path is
multi-threaded like OpenDaylight's, but avoids the YANG datastore round-trip, so it
sits between Ryu's direct \texttt{FlowMod} path and OpenDaylight's model-driven one
in software depth.

The three approaches consequently suggest different per-operation costs. Ryu's
direct state and flow-programming path requires the least memory and software
overhead but serializes on one core. The two JVM controllers can overlap independent
control-plane operations; between them, ONOS's lighter in-memory flow path is
expected to consume less CPU and heap than OpenDaylight's model-driven datastore.

These architectural differences motivate our measurements of reactive
latency, flow-setup throughput, session continuity, and controller resource
consumption in Sections~\ref{sec:methodology}--\ref{sec:discussion}.






\section{Evaluation Methodology}
\label{sec:methodology}

We compare CPAM running on Ryu, OpenDaylight, and ONOS while holding the MTD
logic, topology, workloads, configuration, and measurement procedure
constant. The evaluation follows metrics aligned with the RFC~8456
controller-benchmarking methodology~\cite{rfc8456}, extended with
rotation-aware latency and session-continuity measurements for
address-shuffling MTD.

\subsection{Common CPAM Implementation}
\label{subsec:implementation}

All three controllers implement CPAM~\cite{chakkour2026balancing}. Each protected host
has a stable real IP address and a rotating VIP exposed to peers. Every 60\,s,
the controller assigns new primary VIPs and atomically publishes the updated
mapping. Translation occurs at the destination access switch, so packets cross
the fabric using VIPs before last-hop translation.

A bidirectional session binding pins the VIPs used when a connection is
established. New sessions use current primary VIPs, while established sessions
retain their pinned mappings across rotations. Retired VIPs and session-scoped
rules are reclaimed only after their session and flow references expire. A
packet without a matching rule triggers a \texttt{PacketIn}; the controller
resolves the VIP, computes the path, and installs forwarding and translation
rules. Algorithm~\ref{alg:cpam} summarizes the common control logic, while
Table~\ref{tab:shared-configuration} lists the shared configuration. Mapping
publication is atomic, preventing traffic generators from observing a
partially updated mapping.

\begin{algorithm}[t]
\caption{CPAM rotation and reactive-processing logic.}
\label{alg:cpam}
\small
\begin{algorithmic}[1]
\Statex \textbf{State:} $\mathrm{primary}[h]$: current VIP of host $h$;
  $\mathrm{SBT}$: bidirectional session bindings, keyed by flow; each VIP in state
  \textsc{Primary}, \textsc{Grace}, or \textsc{Quarantine}.
\Statex
\Procedure{OnRotationTick}{ } \Comment{every rotation interval (60\,s)}
  \ForAll{protected hosts $h$}
    \State $v_{\mathrm{old}} \gets \mathrm{primary}[h]$;\ \ $v_{\mathrm{new}} \gets$ \Call{AllocVIP}{ }
    \State $\mathrm{primary}[h] \gets v_{\mathrm{new}}$;\ mark $v_{\mathrm{new}}$ \textsc{Primary}
    \State mark $v_{\mathrm{old}}$ \textsc{Grace} \Comment{still valid for pinned sessions}
  \EndFor
  \State publish the virtual-to-real mapping \emph{atomically}
  \ForAll{VIPs $v$ in state \textsc{Grace}}
    \If{$\mathrm{flowRefs}(v){=}0 \wedge \mathrm{sessionRefs}(v){=}0$ beyond the reclaim threshold}
      \State delete $v$'s forwarding rules;\ mark $v$ \textsc{Quarantine}
    \EndIf
  \EndFor
  \State return quarantined VIPs to the pool after the quarantine period
\EndProcedure
\Statex
\Procedure{OnPacketIn}{$pkt$} \Comment{packet with no matching flow}
  \State classify $pkt$ by source/destination namespace (real vs.\ VIP)
  \State $f \gets$ normalized flow key of $pkt$
  \If{$f \notin \mathrm{SBT}$} \Comment{new protected session}
    \State $v_c \gets \mathrm{primary}[\mathrm{src}]$;\ \ $v_s \gets$ current VIP of the destination
    \State $\mathrm{SBT}[f], \mathrm{SBT}[\bar{f}] \gets$ binding that pins $(v_c, v_s)$
    \State increment session and flow references on $v_c, v_s$
  \Else
    \State reuse the pinned binding $\mathrm{SBT}[f]$ \Comment{survives rotation}
  \EndIf
  \State compute the forwarding path to the destination access switch
  \State install path flows, with last-hop VIP$\rightarrow$real translation
  \State forward $pkt$
\EndProcedure
\end{algorithmic}
\end{algorithm}

\begin{table}[t]
\centering
\caption{Configuration shared by the Ryu, OpenDaylight, and ONOS
implementations.}
\label{tab:shared-configuration}
\small
\begin{tabular}{p{0.58\columnwidth}p{0.27\columnwidth}}
\toprule
\textbf{Parameter} & \textbf{Value} \\
\midrule
Rotation interval             & 60\,s \\
VIP pool size                 & 6,000 \\
Rotation batch size           & 20 hosts \\
Rotation batch delay          & 150\,ms \\
TCP session idle timeout      & 15\,s \\
UDP active timeout            & 30\,s \\
Grace-state reclaim threshold & 5\,s \\
VIP quarantine period         & 30\,s \\
VIP-rule priority             & 100 \\
Table-miss priority           & 0 \\
OpenFlow version              & 1.3 \\
Translation placement         & Last hop \\
Mapping publication           & Atomic \\
\bottomrule
\end{tabular}
\end{table}

\subsection{Control-Plane Workload and Complexity}
\label{sec:complexity}

We characterize the controller work generated by CPAM. Let $N$
denote the number of protected hosts, $G$ the number of VIPs
currently in the Grace state, $H$ the number of switches on a
forwarding path, and $\lambda$ the arrival rate of new reactive
connections. Let $T$ denote the mutation interval.

During a rotation, the controller updates the primary VIP of every
protected host and examines the Grace-state VIPs for reclamation.
Assuming constant-time VIP allocation and reference-count access,
the controller-side processing complexity of a rotation is
\[
    C_{\mathrm{rot}} = O(N+G).
\]
This bound describes state-processing work; the actual completion
time also depends on the number of generated OpenFlow operations
and how the runtime schedules them.

For a new connection, the controller performs an expected
$O(1)$ session-table lookup, computes a forwarding path, and
installs rules along that path. If $C_{\mathrm{path}}$ denotes the
path-computation cost, the reactive-handler complexity is
\[
    C_{\mathrm{react}} = O(C_{\mathrm{path}}+H).
\]
With cached paths, $C_{\mathrm{path}}=O(1)$ and the reactive cost is
$O(H)$. Without caching, a graph traversal requires
$O(|V_s|+|E_s|)$ time, where $V_s$ and $E_s$ are the switch
vertices and links in the topology.

The resulting asymptotic control-plane work rate is
\[
    \Phi(T,\lambda)
    =
    O\!\left(
        \frac{N+G}{T}
        +
        \lambda(C_{\mathrm{path}}+H)
    \right).
\]
Thus, reducing the mutation interval increases periodic controller
work as $1/T$, while connection churn increases reactive work
linearly with $\lambda$. The runtime determines whether these two
components are serialized or processed concurrently.

In Ryu, the latency of a reactive event that overlaps a non-yielding
rotation segment can be expressed as
\[
    L_{\mathrm{Ryu}}
    =
    L_{\mathrm{base}} + R_{\mathrm{rot}},
\]
where $R_{\mathrm{rot}}$ is the residual rotation work before the
cooperative scheduler yields. For OpenDaylight,
\[
    L_{\mathrm{ODL}}
    =
    L_{\mathrm{base}} + Q_{\mathrm{pool}},
\]
where $Q_{\mathrm{pool}}$ captures worker-pool queuing and
synchronization overhead. ONOS is likewise a thread-pool JVM runtime
and takes the same form,
\[
    L_{\mathrm{ONOS}}
    =
    L_{\mathrm{base}} + Q'_{\mathrm{pool}},
\]
where $Q'_{\mathrm{pool}}$ is the corresponding queuing and coordination
term for its in-memory flow subsystem. Because that path avoids
OpenDaylight's model-driven datastore round-trip, $Q'_{\mathrm{pool}}$
is expected to be smaller than $Q_{\mathrm{pool}}$, consistent with
ONOS's lower measured resource cost at comparable reactive latency
(Section~\ref{sec:results}). The model thus separates the single
cooperative runtime (Ryu), whose delay is dominated by the rotation
residual $R_{\mathrm{rot}}$, from the two thread-pool runtimes, whose
additional delay is instead worker-pool queuing. These expressions do
not assume a particular measured latency; rather, they identify the
scheduling components evaluated in Sections~\ref{sec:results} and
\ref{sec:discussion}.

\subsection{Testbed and Topology}
\label{subsec:testbed}

\begin{figure}[t]
    \centering
    \includegraphics[
        width=\linewidth,
        height=0.5\textheight,
        keepaspectratio
    ]{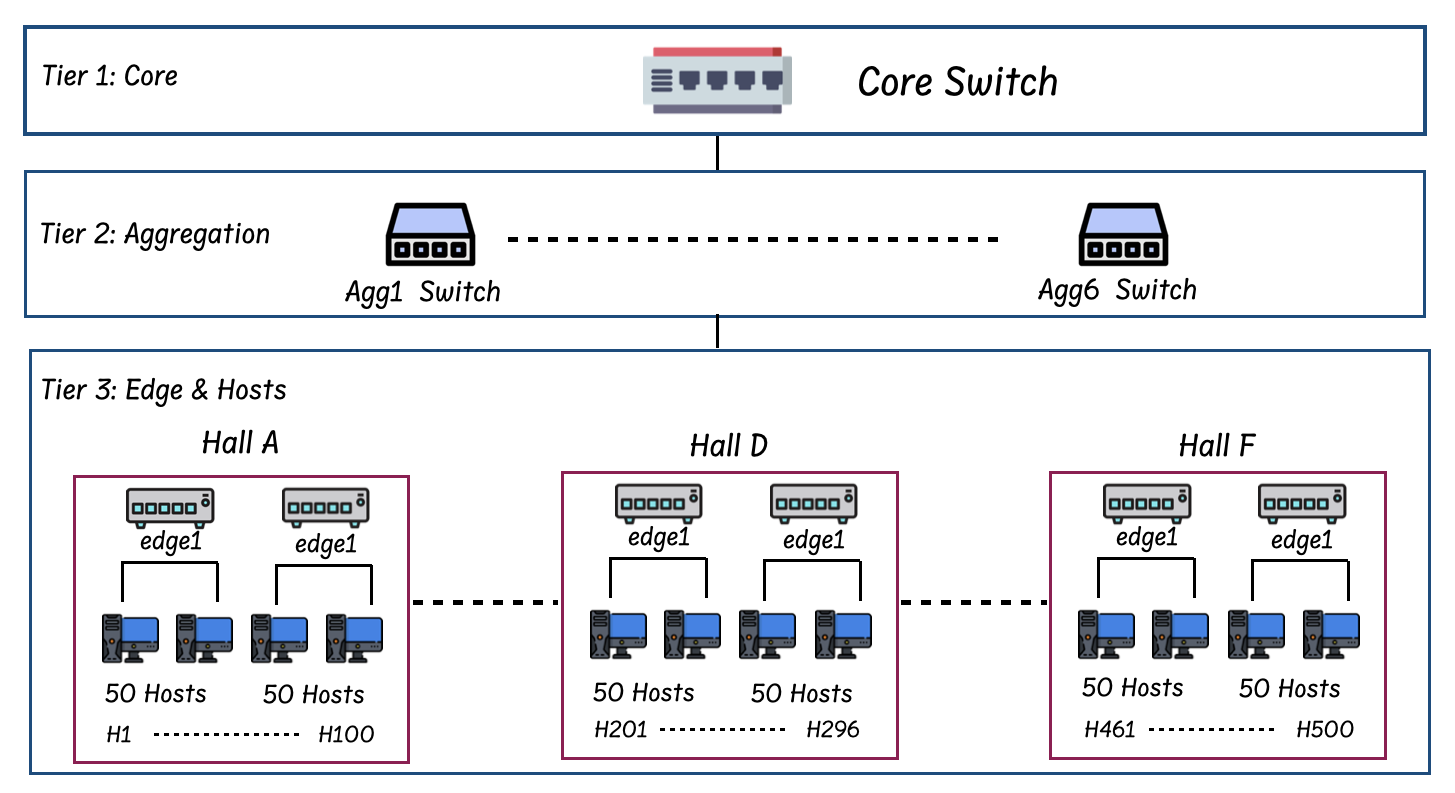}
    \caption{Three-tier evaluation fabric with 19 switches and 500 hosts in
    250 communicating pairs.}
    \label{fig:approach}
\end{figure}

Mininet and Open vSwitch emulate the same three-tier campus fabric for all three
controllers: one core, six aggregation, and twelve edge switches connecting
500 hosts in 250 communicating pairs (Figure~\ref{fig:approach}). Host
assignments and traffic sequences are identical across controllers.

\subsection{Performance Metrics}
\label{subsec:metrics}

We evaluate three aspects of controller behavior:
\emph{data-plane correctness}, \emph{control-plane responsiveness},
and \emph{resource cost}.

\noindent\textbf{Data-plane correctness.}
UDP packet loss and jitter measure whether continuous mutation affects
packet delivery. TCP session continuity measures the fraction of
long-lived sessions that remain operational across at least one
complete address-rotation cycle.

\noindent\textbf{Control-plane responsiveness.}
Reactive round-trip time (RTT) measures the end-to-end latency of a
packet that triggers a \texttt{PacketIn}, including controller
processing, rule installation, and the first reply. Reactive
per-packet processing time (RPPT) isolates controller-side processing
from \texttt{PacketIn} receipt to \texttt{FlowMod} transmission. We
also report flow-setup rate (FSR), measured as the number of new flows
installed per second during connection churn.

\noindent\textbf{Resource cost.}
We measure controller CPU utilization and memory consumption. Because
the JVM may reserve substantially more memory than it actively uses,
the JVM controllers (OpenDaylight and ONOS) report memory using both
resident set size and post-garbage-collection live heap. Ryu memory is
reported as resident set size.

\subsection{Workload Generation}
\label{subsec:workloads}

Each run executes the same sequence of traffic phases on all three
controllers. A settling period separates consecutive phases and
allows forwarding state to stabilize.

\begin{enumerate}[leftmargin=*,nosep]
    \item \textbf{UDP QoS:} Each host pair generates a
    constant-bit-rate UDP stream to measure jitter and packet loss.

    \item \textbf{Reactive ICMP:} Each pair sends ICMP requests to a
    newly selected VIP, forcing reactive rule installation and
    providing the reactive RTT measurement.

    \item \textbf{TCP throughput:} Each pair performs a bulk TCP
    transfer to exercise the forwarding path.

    \item \textbf{Session continuity:} Each pair opens a long-lived
    TCP connection that spans multiple rotation intervals.

    \item \textbf{Connection churn:} Each pair repeatedly creates
    short-lived TCP connections. Each unmatched connection generates
    reactive controller work and is used to measure FSR and RPPT under
    load.
\end{enumerate}

Each connection resolves its destination VIP immediately before
connection establishment using the controller's published mapping.
Resolution uses bounded retry if a rotation occurs during the lookup.
This prevents a workload launched before a rotation from using a stale
VIP and incorrectly attributing the resulting failure to the
controller.

\subsection{Measurement Instrumentation}
\label{subsec:instrumentation}

Reactive RTT, UDP jitter and loss, and TCP behavior are collected from
\texttt{ping} and \texttt{iperf} client logs. RPPT is measured using
controller-generated timestamps around each
\texttt{PacketIn}-to-\texttt{FlowMod} processing interval. This
measurement excludes network-channel and data-plane delay.

FSR and forwarding-table occupancy are derived from periodic
\texttt{ovs-ofctl} snapshots collected during the connection-churn
phase. Controller CPU utilization is sampled with \texttt{pidstat}
and restricted to the controller process. Memory is measured from
process resident size; for the JVM controllers (OpenDaylight and ONOS),
a forced garbage collection followed by a heap query additionally
provides the live JVM heap.





\section{Results}\label{sec:results}

Table~\ref{tab:main} summarizes the ten-run results. Data-plane
correctness is similar across controllers, whereas control-plane latency and
resource use differ sharply.

\subsection{Reactive Round-Trip Time}

\begin{figure}[H]
  \centering
  \begin{subfigure}[t]{0.49\columnwidth}
    \includegraphics[width=\linewidth]{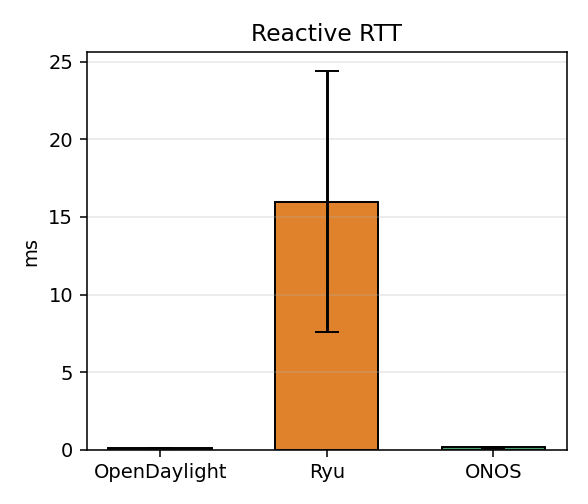}
    \caption{Mean $\pm$ std}
  \end{subfigure}\hfill
  \begin{subfigure}[t]{0.49\columnwidth}
    \includegraphics[width=\linewidth]{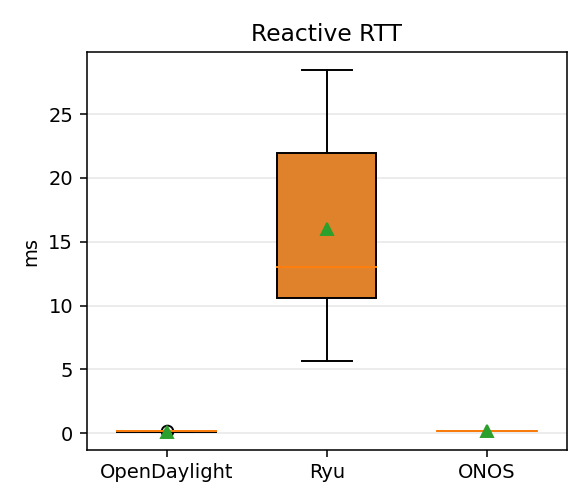}
    \caption{Per-run distribution}
  \end{subfigure}
  \caption{Reactive RTT across ten runs. OpenDaylight and ONOS both remain tightly
clustered and sub-millisecond, whereas Ryu exhibits a substantially larger mean and
run-to-run spread.}
  \label{fig:rtt}
\end{figure}

Reactive RTT is the user-visible cost of the control plane: the round-trip time of a
connection whose first packet must trigger a flow installation before it can be
forwarded. It is $0.131\pm0.014$\,ms on OpenDaylight and $0.159\pm0.014$\,ms on ONOS,
against $15.98\pm8.41$\,ms on Ryu (Figure~\ref{fig:rtt}). The two JVM controllers land
within $0.03$\,ms of each other---roughly two orders of magnitude below Ryu.
Ryu's minimum remains sub-millisecond, but its run-level values are much
more widely distributed. Section~\ref{sec:discussion} relates this variation
to the timing of reactive events relative to address rotation.

\subsection{Reactive Processing Time}

\begin{figure}[t]
  \centering
  \includegraphics[width=\columnwidth]{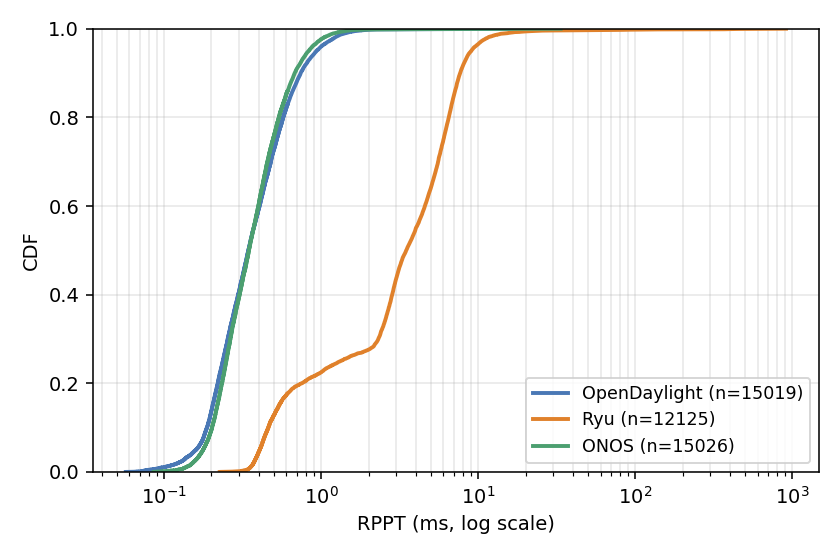}
  \caption{Cumulative distribution of RPPT, pooled over the ten-run set (log-scale
  $x$). OpenDaylight and ONOS are concentrated below $2$\,ms; Ryu is shifted an order of
  magnitude higher with a tail extending to $\sim$$1$\,second, the same shape at idle
  and under churn load.}
  \label{fig:rppt_cdf}
\end{figure}

Reactive per-packet processing time (RPPT) isolates the controller-side component of
the reactive path---the interval from \texttt{PacketIn} receipt to \texttt{FlowMod}
departure---and therefore removes channel and data-plane latency. The two JVM
controllers cluster an order of magnitude below Ryu in the mean ($0.44$ and $0.42$
vs $4.86$\,ms), but the mean understates the effect; the full distributions
(Figure~\ref{fig:rppt_cdf}, percentiles in Table~\ref{tab:rppt}) reveal a divergence
in \emph{shape}. Pooling all per-event RPPT samples across the ten runs, OpenDaylight
and ONOS are both tight from median to tail (p50 to p99 spans $0.34$--$1.4$ and
$0.35$--$1.2$\,ms respectively, a factor of about four), whereas Ryu is heavy-tailed:
a $3.5$\,ms median but a p99 of $15.4$\,ms and a maximum of $909$\,ms, a single
installation stalled for nearly one second behind a rotation burst. A concurrent
connection-churn load (``under load'' in Table~\ref{tab:rppt}) barely moves the JVM
controllers (p99 $1.4\!\to\!2.0$\,ms for OpenDaylight, $1.2\!\to\!2.1$\,ms for ONOS),
confirming that their thread pools absorb the additional \texttt{PacketIn} rate,
whereas Ryu's tail rises further (p99 $15.4\!\to\!20.9$\,ms, maximum $1084$\,ms). The
heavy tail, not the median, is what a latency-sensitive application experiences during
rotation.

\begin{table}[t]
\centering
\caption{RPPT percentiles (ms), pooled over all per-event samples in the ten-run
set. Ryu's p99$\gg$p50 is the signature of installations serialized behind the
rotation burst.}
\label{tab:rppt}
\small
\begin{tabular}{@{}llrrrr@{}}
\toprule
\textbf{Controller} & \textbf{Load} & \textbf{p50} & \textbf{p90} & \textbf{p95} & \textbf{p99} \\
\midrule
OpenDaylight & idle       & $0.34$ & $0.74$ & $0.93$ & $1.43$ \\
OpenDaylight & churn load & $0.32$ & $0.72$ & $0.91$ & $1.97$ \\
ONOS         & idle       & $0.35$ & $0.68$ & $0.83$ & $1.22$ \\
ONOS         & churn load & $0.32$ & $0.74$ & $1.01$ & $2.09$ \\
Ryu          & idle       & $3.48$ & $7.66$ & $9.03$ & $15.39$ \\
Ryu          & churn load & $3.86$ & $7.35$ & $11.83$ & $20.92$ \\
\bottomrule
\end{tabular}
\end{table}

\subsection{Flow-Setup Throughput}

\begin{figure}[t]
  \centering
  \begin{subfigure}[t]{0.49\columnwidth}
    \includegraphics[width=\linewidth]{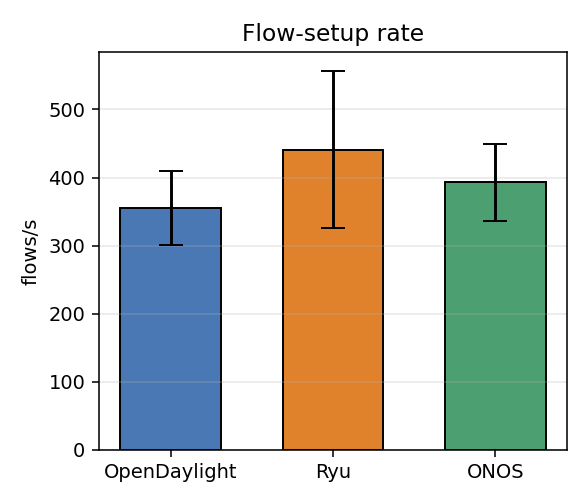}
    \caption{Mean $\pm$ std}
  \end{subfigure}\hfill
  \begin{subfigure}[t]{0.49\columnwidth}
    \includegraphics[width=\linewidth]{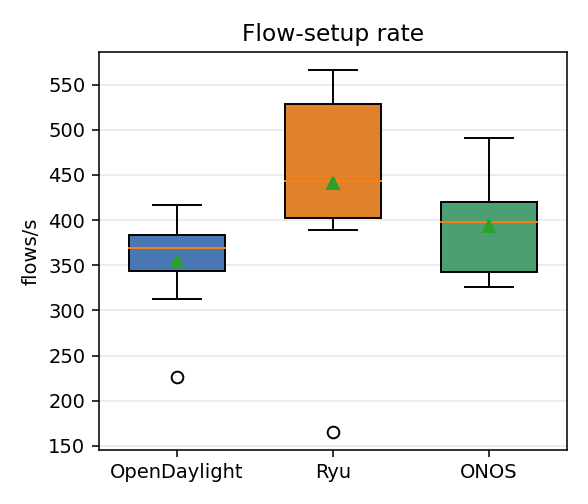}
    \caption{Per-run distribution}
  \end{subfigure}
  \caption{Flow-setup rate under saturating churn across the three controllers. Ryu's
  thinner flow-programming path yields the highest aggregate throughput despite its
  worse per-installation latency.}
  \label{fig:fsr}
\end{figure}

Flow-setup rate (FSR) measures aggregate rule-installation throughput under
the connection-churn workload, reported here as the net Open vSwitch flow-count
delta so that the metric is identical across controllers. Ryu achieves
$441\pm115$ flows/s, ONOS $393\pm56$, and OpenDaylight $355\pm54$
(Figure~\ref{fig:fsr})---all within a comparable band. Thus, the controller with
the highest reactive latency (Ryu) nevertheless provides the greatest aggregate
flow-setup throughput.

This result highlights the distinction between throughput and latency. FSR
measures the total number of installations completed per second under
sustained load, whereas RTT and RPPT measure the delay experienced by
individual reactive events. Consequently, aggregate throughput and
per-installation latency favor different controllers. Section~\ref{sec:discussion}
relates this difference to their flow-programming paths and concurrency
models.

\subsection{Forwarding-Table Occupancy}

\begin{figure}[H]
  \centering
  \begin{subfigure}[t]{0.49\columnwidth}
    \includegraphics[width=\linewidth]{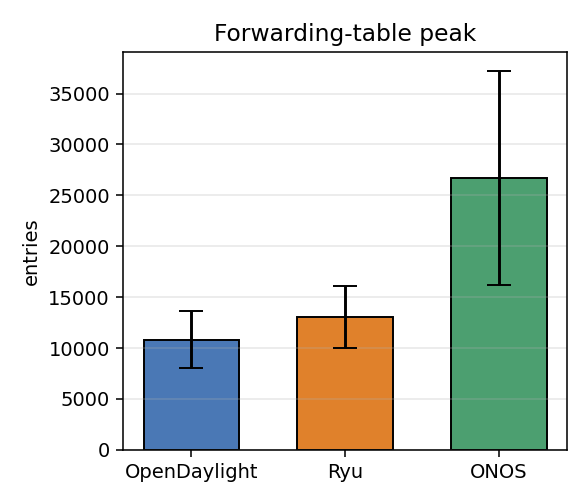}
    \caption{Mean $\pm$ std}
  \end{subfigure}\hfill
  \begin{subfigure}[t]{0.49\columnwidth}
    \includegraphics[width=\linewidth]{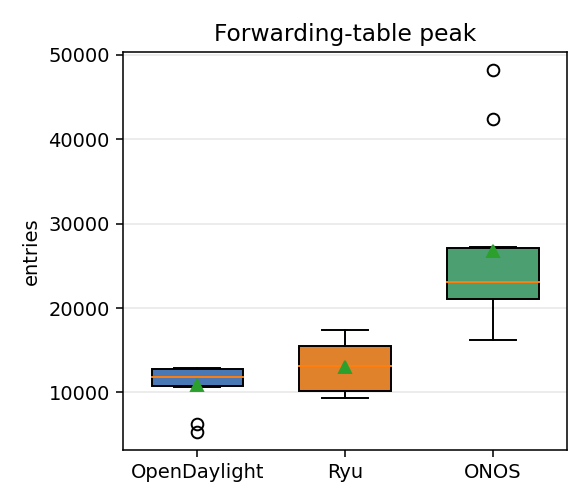}
    \caption{Per-run distribution}
  \end{subfigure}
  \caption{Peak forwarding-table occupancy during churn. The per-session footprint is
  identical by construction; the peak-count differences reflect host-bound
  churn-completion timing (cf.\ TCP throughput), not controller structure.}
  \label{fig:ftable}
\end{figure}

Peak forwarding-table occupancy measures the transient data-plane state induced
during the connection-churn burst. Because every controller runs identical CPAM
logic, the per-session forwarding state is the same by construction; the differences
in peak count---$10\,829\pm2\,793$ entries for OpenDaylight, $13\,048\pm3\,070$ for
Ryu, and $26\,717\pm10\,479$ for ONOS (Figure~\ref{fig:ftable})---therefore reflect
how many concurrent sessions' flows co-reside at the peak instant, which scales with
how quickly the host datapath completes the churn burst. This is the same host-bound
effect observed for TCP throughput (\S\ref{subsec:dataplane}), not a structural
divergence in switch state. Occupancy stays well within a commodity switch's
flow-table capacity at 500 hosts.

\subsection{Controller Resource Overhead}

\begin{figure}[H]
  \centering
  \begin{subfigure}[t]{0.49\columnwidth}
    \includegraphics[width=\linewidth]{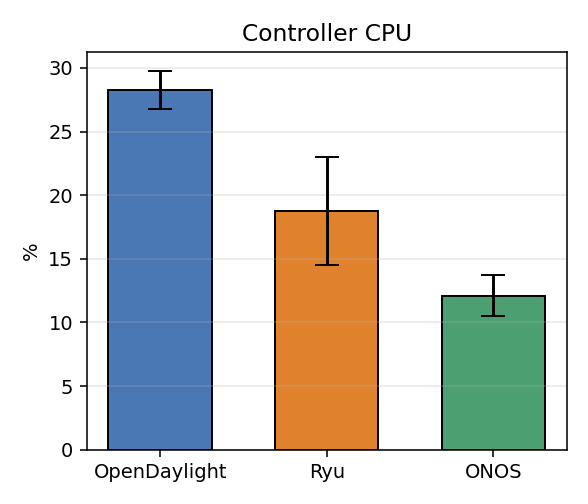}
    \caption{Controller CPU}
  \end{subfigure}\hfill
  \begin{subfigure}[t]{0.49\columnwidth}
    \includegraphics[width=\linewidth]{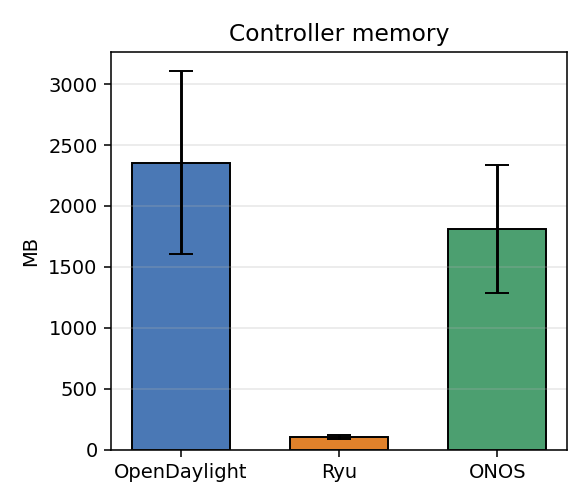}
    \caption{Controller memory}
  \end{subfigure}
  \caption{Controller resource overhead across Ryu, OpenDaylight, and ONOS (mean
  $\pm$ std over 10 runs). Memory is shown as committed resident size; the fair
  post-garbage-collection live heap is $\approx$1\,GB for OpenDaylight and
  $\approx$0.4\,GB for ONOS.}
  \label{fig:overhead}
\end{figure}

The low reactive latency of the JVM controllers comes at a resource cost, though the
two JVM runtimes differ sharply. Average CPU utilization is highest on OpenDaylight
($28.2\pm1.5\%$), lowest on ONOS ($12.1\pm1.6\%$), with Ryu in between ($18.8\pm4.3\%$,
confined to a single core). Memory tracks the runtime: OpenDaylight commits
$\sim$$2.4$\,GB of resident size (fair post-garbage-collection live heap $\sim$$1$\,GB)
and ONOS $\sim$$1.8$\,GB ($\sim$$0.4$\,GB live heap), both an order of magnitude or
more above Ryu's $107\pm17$\,MB (Figure~\ref{fig:overhead}). ONOS thus attains
OpenDaylight-class reactive latency at markedly lower CPU and a smaller live heap,
while Ryu remains by far the most memory-frugal at the cost of its reactive-path
latency and tail.

\subsection{Data-Plane Correctness}
\label{subsec:dataplane}

\begin{figure}[H]
  \centering
  \begin{subfigure}[H]{0.49\columnwidth}
    \includegraphics[width=\linewidth]{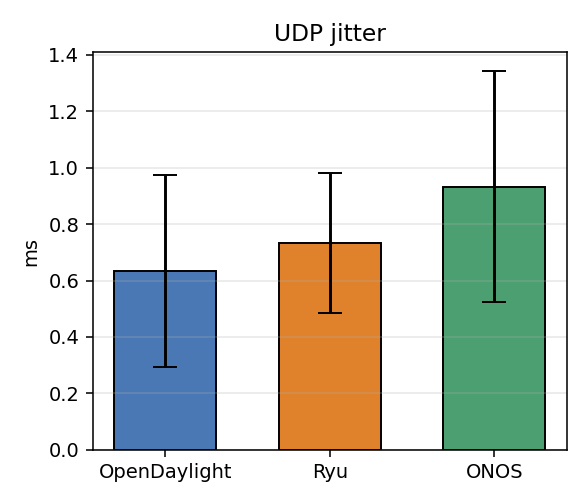}
    \caption{UDP jitter}
  \end{subfigure}\hfill
  \begin{subfigure}[H]{0.49\columnwidth}
    \includegraphics[width=\linewidth]{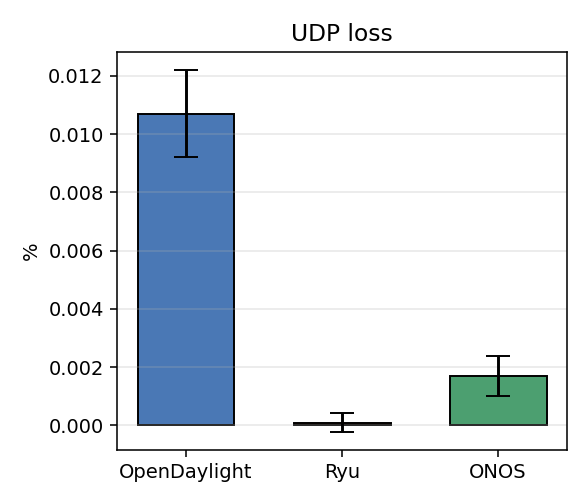}
    \caption{UDP loss}
  \end{subfigure}
  \caption{Data-plane correctness across the three controllers (mean $\pm$ std over
  10 runs). UDP loss and jitter are near-identical, confirming an identical data
  plane. TCP throughput is host-datapath bound (uncapped emulation links) rather than
  controller-distinguishing and is reported in the text (\S\ref{subsec:dataplane}).}
  \label{fig:udp}
\end{figure}

Once flows are installed, all three controllers forward identically. UDP loss is
near-zero everywhere ($\le0.011\%$) and mean jitter is sub-millisecond
($0.63$, $0.73$, and $0.93$\,ms for OpenDaylight, Ryu, and ONOS), as shown in
Figure~\ref{fig:udp}. This is expected: the per-packet translation and forwarding
rules are identical, and every controller programs the same Open vSwitch datapath.
Bulk TCP throughput is bounded by the host software datapath rather than by the
controller---once a flow is installed the controller is no longer on the packet path,
and the emulation links are uncapped---so it is not a controller-distinguishing
metric; measured in a single session on the same host, all three forward at the same
rate ($617$--$673$\,Mbit/s). The data plane therefore does not distinguish the
runtimes; the entire difference between them is in the \emph{control} plane.

\subsection{Session Continuity}

All three controllers keep near-all long-lived sessions alive across rotations,
and---the key qualitative result---\textbf{no controller ever tore down an
established, data-carrying session}; every observed loss occurs at connection
\emph{setup} or is an emulator-saturation event (\S\ref{sec:analysis}). Over ten
back-to-back continuity runs each, OpenDaylight and ONOS each preserve $100\%$ of
sessions and Ryu preserves $99.9\%$ ($3$ of $2500$ sessions broke, range
$99.2$--$100\%$). Each Ryu loss was a
setup-time race with a rotation---two mid-establishment resets within the first
sub-second and one connection that never established. When continuity is instead
measured within the full mixed-traffic workload, where a burst of $250$ simultaneous
connection setups competes with concurrent UDP, ICMP, and TCP traffic, Ryu's
continuity decreases to $96.8\%$, again entirely setup-time contention rather than
mid-session teardown. Continuity is thus best read as a \emph{mechanism} result
(\S\ref{sec:analysis}): the aggregate percentages are near-identical, and the
substantive finding is that established sessions are never broken by a rotation on
any of the three controllers.

\begin{table}[H]
\centering
\caption{Results on the 500-host fabric (mean$\,\pm\,$sd over ten runs). The JVM
controllers' memory gives committed RSS with fair post-GC live heap in parentheses.
$\ddagger$: dedicated continuity workload. $\dagger$: TCP throughput is bounded by the
host software datapath (uncapped emulation links), not the controller; values are a
same-session measurement (\S\ref{subsec:dataplane}). $\S$: peak concurrent flow
count during churn; the per-session footprint is identical by construction, so the
difference reflects host-bound churn-completion timing (cf.\ TCP), not controller structure.}
\label{tab:main}
\resizebox{\columnwidth}{!}{%
\begin{tabular}{@{}lrrr@{}}
\toprule
\textbf{Metric} & \textbf{OpenDaylight} & \textbf{Ryu} & \textbf{ONOS} \\
\midrule
\multicolumn{4}{@{}l}{\emph{Data plane}}\\
UDP loss (\%) & $0.011\pm0.001$ & $0.000\pm0.000$ & $0.002\pm0.001$ \\
UDP jitter (ms) & $0.63\pm0.32$ & $0.73\pm0.24$ & $0.93\pm0.41$ \\
TCP throughput (Mbit/s)$^{\dagger}$ & $617$ & $639$ & $673$ \\
\midrule
\multicolumn{4}{@{}l}{\emph{Control plane}}\\
Reactive RTT (ms) & $0.131\pm0.014$ & $15.98\pm8.41$ & $0.159\pm0.014$ \\
RPPT mean (ms) & $0.44\pm0.06$ & $4.86\pm1.36$ & $0.42\pm0.04$ \\
Session continuity (\%) & $100.0^{\ddagger}$ & $99.9$ & $100.0^{\ddagger}$ \\
Flow-setup rate (fl/s) & $355\pm54$ & $441\pm115$ & $393\pm56$ \\
Forwarding-table peak$^{\S}$ & $10\,829\pm2\,793$ & $13\,048\pm3\,070$ & $26\,717\pm10\,479$ \\
\midrule
\multicolumn{4}{@{}l}{\emph{Resource cost}}\\
Controller CPU (\%) & $28.2\pm1.5$ & $18.8\pm4.3$ & $12.1\pm1.6$ \\
Memory & $2.4$ GB ($\approx1$ GB) & $107\pm17$ MB & $1.8$ GB ($\approx0.4$ GB) \\
\bottomrule
\end{tabular}}
\end{table}

\section{Analysis and Discussion}
\label{sec:analysis}

The results reveal an observed resource-versus-responsiveness trade-off.
Because the CPAM mechanism, topology, workload, configuration, and
measurement procedure are controlled across the three implementations, the
differences are consistent with the controllers' execution, state-management,
and flow-programming architectures. We interpret these results using the
control-plane workload model in Section~\ref{sec:complexity}.

\subsection{Runtime-Level Mechanisms}
\label{subsec:root-causes}

\noindent\textbf{Rotation scheduling.}
Ryu multiplexes address rotation, \texttt{PacketIn} handling, and statistics
collection through a cooperative event loop. A reactive event arriving during
a non-yielding rotation segment cannot be processed until that segment yields.
It therefore incurs the residual rotation delay $R_{\mathrm{rot}}$ identified
in Section~\ref{sec:complexity}. This behavior is consistent with the RPPT
distribution in Figure~\ref{fig:rppt_cdf}: Ryu remains fast outside rotation
epochs, but events overlapping rotation form a substantially heavier tail.

OpenDaylight distributes rotation-related work and reactive OpenFlow
processing across worker and service threads. Reactive processing can
therefore proceed without waiting for the active rotation segment to finish.
Its additional delay is instead associated with worker-pool queuing and
synchronization, represented by $Q_{\mathrm{pool}}$ in
Section~\ref{sec:complexity}. This execution model is consistent with
OpenDaylight's tighter reactive-latency distribution.

ONOS provides an independent confirmation of this mechanism. As a second,
architecturally distinct JVM controller, it dispatches reactive \texttt{PacketIn}
processing on a thread pool separate from its rotation and statistics timers, and it
reproduces the same low, flat reactive-latency distribution as OpenDaylight
(Table~\ref{tab:rppt}). That two unrelated JVM runtimes both avoid the heavy tail
indicates the effect stems from the concurrency model---parallel dispatch of reactive
and rotation work---rather than any implementation detail specific to OpenDaylight.
Where the two JVM controllers diverge is resource cost: ONOS programs flows through an
in-memory \texttt{FlowRuleService}, whereas OpenDaylight serializes each installation
through its MD-SAL datastore, and this heavier per-operation path is consistent with
OpenDaylight's larger CPU utilization and live heap at equal reactive latency.

As a diagnostic check, explicitly yielding inside the Ryu rotation loop using
\texttt{hub.sleep(0)} substantially reduced its reactive tail. This observation
further supports cooperative serialization, rather than address translation
itself, as the principal source of rotation-correlated delay.

\noindent\textbf{Setup-time races.}
Once CPAM creates a session binding and installs its forwarding state, the
session remains pinned to the VIPs under which it was established. Those VIPs
and their associated rules are retained until their session and flow
references expire. Consequently, a subsequent rotation does not invalidate
the forwarding state of an established session, consistent with the absence
of mid-session failures in Section~\ref{sec:results}.

The observed failures instead occur before a connection becomes a stable,
data-carrying session. A new connection may resolve or begin using a VIP while
the corresponding mapping and forwarding state are being updated. Depending
on the ordering of these operations, its initial packets may be delayed,
reset, or fail to reach the destination.

The duration of this vulnerable setup window depends partly on controller
scheduling. Ryu's serialized execution can keep reactive processing waiting
while rotation work remains active, increasing the opportunity for overlap.
OpenDaylight can process reactive events concurrently with rotation, reducing
that window. The relevant distinction is therefore between
\emph{setup-time reliability} and \emph{preservation of established
sessions}.

\noindent\textbf{Throughput and resources.}
Ryu's direct flow-programming path and lightweight single-process runtime
impose relatively little per-flow framework overhead. When the rotation
handler is inactive, this path can complete many rule installations, which is
consistent with its higher aggregate flow-setup rate. The same cooperative
execution model, however, cannot overlap a reactive installation with a
non-yielding rotation segment.

OpenDaylight incurs additional coordination through its worker pools,
MD-SAL services, managed state, and JVM runtime. These components increase
CPU and memory consumption and introduce per-flow processing overhead, but
they also isolate reactive work from periodic mutation processing.
Consequently, higher aggregate throughput and lower per-event latency need
not belong to the same controller: FSR measures completed installations per
unit time, whereas RTT and RPPT measure the delay experienced by individual
events.

\subsection{Implications for MTD Design}
\label{sec:discussion}

\noindent\textbf{Joint configuration.}
The workload model in Section~\ref{sec:complexity} shows that periodic
controller work increases as the mutation interval $T$ decreases, while
reactive work increases with the new-connection rate $\lambda$. A controller
that performs well under infrequent rotations may therefore behave
differently when rotations become more frequent or connection arrivals become
more bursty.

Under the tested configuration, the three controllers trace out distinct
operating points rather than a single trade-off curve. Ryu is attractive when
controller resources are constrained and occasional reactive tail latency is
acceptable. The JVM controllers provide stronger isolation between rotation and
connection establishment---lower, flatter reactive latency and better setup-time
reliability---at higher resource cost; but they are not equivalent. ONOS reaches
this low-latency operating point at markedly lower CPU and a smaller live heap
than OpenDaylight, making it the more resource-efficient choice when JVM-class
responsiveness is required, whereas OpenDaylight's model-driven services incur the
highest resource cost of the three. These results do not establish one platform as
universally superior; they identify three operating points---resource-frugal but
tail-prone (Ryu), responsive but heavy (OpenDaylight), and responsive-and-leaner
(ONOS)---whose suitability depends on deployment requirements.

\noindent\textbf{Evaluation metrics.}
Aggregate throughput alone does not capture the delay experienced by
individual connections. Likewise, an aggregate continuity percentage does
not distinguish a connection that fails during setup from an established
session disrupted by rotation. Evaluations of SDN-based MTD should therefore
report tail-latency metrics such as p95 and p99 RPPT, separate setup-time
failures from mid-session failures, and report both throughput and per-event
latency.

Resource measurements should also account for runtime architecture. For a
JVM-based controller, resident set size and post-garbage-collection live heap
describe different aspects of memory use and should be reported separately
when comparing against a lightweight controller runtime.

\noindent\textbf{Incremental rotation.}
The Ryu results do not imply that cooperative controllers are inherently
unsuitable for address mutation. Their blocking interval can be shortened by
dividing rotation work into smaller batches, yielding between batches,
incrementally publishing updates, or proactively installing forwarding state
for the next VIP assignment. These techniques reduce the residual rotation
work faced by an arriving reactive event.

Thread-pooled controllers may instead benefit from tuning worker pools, JVM
memory, datastore transactions, and flow-programming services. More generally,
high-frequency mutation should avoid a monolithic run-to-completion update and
be implemented incrementally so that reactive connection setup can continue.

\subsection{Limitations}
\label{sec:limitations}

The absolute measurements are specific to the three tested controller platforms,
the $500$-host emulated topology, and the fixed $60$\,s mutation interval.
The controllers were also evaluated through back-to-back runs without a
restart between runs, and all $19$ switches shared one software datapath.
Sustained execution in this environment may increase run-to-run variance and
does not reproduce the behavior of physical or distributed switching
hardware.

The matched experimental configuration supports the relative comparison
reported in this paper, but the measured values should not be generalized to
all controllers, mutation frequencies, workloads, or deployment platforms.
Restart-isolated experiments, additional controllers, varied mutation
intervals, and physical or distributed testbeds are left for future work.

\section{Conclusion and Future Work}
\label{sec:conclusion}

This paper shows that the SDN controller platform materially affects
the performance of address-shuffling Moving Target Defense. By
implementing identical CPAM logic on Ryu, OpenDaylight, and ONOS and
evaluating all three on the same $500$-host topology, we isolate the
effect of controller architecture. All three implementations preserve
data-plane correctness and established sessions across address
rotations, confirming that CPAM's continuity mechanism operates
consistently across the platforms.

The controllers nevertheless expose different
resource-versus-responsiveness operating points. The two JVM
controllers achieve substantially lower reactive latency and stronger
connection-establishment reliability during rotations, whereas Ryu
uses considerably less memory and provides higher aggregate
flow-setup throughput. The observed differences are consistent with
their execution models: Ryu serializes rotation and reactive work
through cooperative scheduling, while the JVM controllers process them
concurrently through worker threads. Crucially, the two JVM runtimes
are not interchangeable: ONOS attains OpenDaylight-class reactive
latency at the lowest CPU utilization of the three and a smaller live
heap than OpenDaylight, tracing to its lighter in-memory flow path
versus OpenDaylight's model-driven datastore---low reactive latency
therefore need not carry OpenDaylight's full resource cost. Controller
selection should be considered together with mutation frequency,
workload intensity, latency requirements, and available resources.

Our evaluation is limited to three controllers, one emulated topology,
a fixed $60$\,s mutation interval, and back-to-back runs on a shared
software datapath. Future work will validate the results using
restart-isolated runs and physical or distributed testbeds, and vary
the mutation interval and workload intensity. We will also explore
incremental rotation, proactive rule installation, and programmable
data planes such as P4~\cite{bosshart2014p4} and
eBPF~\cite{hoiland2018xdp,cilium} to reduce control-plane contention
and support higher-frequency mutation.

\bibliographystyle{unsrtnat}
\bibliography{references}

\end{document}